\documentclass[aps,prb,twocolumn,floatfix,showpacs,amssymb]{revtex4}

\usepackage{indentfirst}
\usepackage{amsmath}
\usepackage{graphicx}

\usepackage{amsfonts}

\usepackage{amssymb}

\begin{document}

\title{ { {Critical parameters for the one-dimensional systems with long-range correlated disorder }}}

\author{Yi Zhao, Suqing Duan, and Wei Zhang {\footnote{Author to whom any correspondence should be addressed, Email:
zhang$\_$wei@iapcm.ac.cn}}}

\affiliation{Institute of Applied Physics and Computational
Mathematics, P.O. Box 8009(28), Beijing 100088, P. R. China}

\begin{abstract}
\smallskip
We study the metal-insulator transition in a tight-binding
one-dimensional (1D) model with long-range correlated disorder. In
the case of diagonal disorder with site energy within
$[-\frac{W}{2},\frac{W}{2}]$ and having a power-law spectral density
$S(k)\propto k^{-\alpha}$, we investigate the competition between
the disorder and correlation. Using the transfer-matrix method and
finite-size scaling analysis, we find out that there is a finite
range of extended eigenstates for $\alpha>2$, and the mobility edges
are at $\pm E_{c}=\pm|2-W/2|$. Furthermore, we find the critical
exponent $\nu$ of localization length ($\xi \sim |E-E_{c}|^{-\nu}$)
to be $\nu=1+1.4e^{2-\alpha}$. Thus our results indicate that the
disorder strength $W$ determines the mobility edges and the degree
of correlation $\alpha$ determines the critical exponents.
\end{abstract}

 \maketitle
\setcounter{section}{0} \setcounter{equation}{0}
\setcounter{subsection}{0} \setcounter{subsubsection}{0}

\section{Introduction}
Anderson's localization theory points out that in a system with
uncorrelated diagonal disorder, all one-electron states are
spatially localized with an exponentially decaying envelope with a
characteristic localization length $\xi$, when the disorder strength
is larger than a critical value.$\cite{1}$ From the scaling
theory,$\cite{2}$ it is also well-known that an infinitesimal
disorder can cause localization of all states in one and two
dimensional systems in the thermodynamic limit. In recent years, it
was found that spatial correlation of disorder potential played an
important role in the nature of long-range charge transport of
low-dimensional systems. In the presence of short-range correlated
disorder, there exists a set of discrete resonant energy levels of
extended states.$\cite{3,4,5,6,7,8,9,10,11,12}$ For example, a
single extended eigenstate was found in the random dimer
model,$\cite{4,8,12}$ which was verified by measuring microwave
transmission on the semiconductor superlattices with intentional
correlated disorder.$\cite{13}$ These models with short-range
correlated disorder do not possess a true disorder induced
metal-insulator transition in the thermodynamic limit. More
recently, a one-dimensional (1D) disordered model with long-range
correlation considered by de Moura and Lyra has arisen a great
interest, as it can result in a continue band of extended states
under appropriate conditions.$\cite{14}$ The transition from
localization to delocalization occurred at the critical points was
examined experimentally in a single-mode waveguide with inserted
correlated scatters.$\cite{15}$ More recent experiments in
ultra-cold atoms/Bose-Einstein condensate push the research in this
direction further. \cite{bec}

In the disordered systems, the critical parameters, such as the
critical disorder strength $W_c$, mobility edge $E_c$, critical
exponents $\nu_E$, $\nu_W$ (defined by the dependence of
localization length on $W$ or $E$ near the critical points $\xi\sim
|W-W_c|^{\nu_W}$ or $\xi\sim |E-E_c|^{\nu_E}$) are of particular
importance and interest, since they determine the nature of the
localization-delocalization transitions and the universal properties
of disordered systems. Even for three-dimensional (3D) Anderson
model which has been studied for more than forty years, the accurate
determination of the critical parameters is still under recent
investigations,$\cite{16}$ where the critical parameters for
different lattices are calculated and it was found that $\nu_E$ is
close to $\nu_W$. While for 3D systems with scale-free disorder, it
was found that $\nu_E \neq \nu_W$ and the extended Harris criterion
is obeyed. $\cite{16b}$

In spite of many
studies$\cite{3,4,5,6,7,8,9,10,11,12,13,14,15,17,18,19,20}$ on the
1D systems with correlated disorder, many of them are in a
perturbative level and  there are still several important questions
to be answered, such as the positions of the mobility edges and the
value of the critical exponent $\nu$. The determination of these
critical parameters and the comparison with the results of 3D models
mentioned above will definitely deepen our understanding on
disordered systems, especially the dependence on dimensionality and
correlation. Moreover, the correlated disordered system is a kind of
system which is intermediate between ordered system and pure
disordered system. The investigation of the universal properties of
this type system  and comparison with quasi-periodic system is of
fundamental importance. In addition, the accurate values of these
critical parameters (mobility edges) have important applications in
charge transport in low dimensional (correlated) disordered systems,
such DNA molecules,$\cite{21}$, Bose-Einstein condensate in 1-D
optical lattices with speckle potential.\cite{bec}. In the present
paper, we calculate these critical parameters for the 1D systems
with the correlated diagonal disorder. And we also resolve some
inconsistencies in the literatures.$\cite{14}$ One of the advantages
of our study is that we are able to investigate
localization-delocalization transition and obtain the accurate
critical parameters in 1D system with large size and the boundary
effects is negligible, unlike the 3D system where all the sizes in
three dimensions can not be very large due to the computational
limitation.

 We first investigate the
localization properties of the 1D tight-binding model with
long-range correlated diagonal disorder (the hopping constant is set
to be unity). The diagonal on-site energies $\{\epsilon_{i}\}$ are
distributed in $[-W/2, W/2]$ with $W$ being the strength of
disorder. The site energies have a power-law spectral density
$S(k)\propto k^{-\alpha}$ with $\alpha>0$. The function $S(k)$ is
the Fourier transform of the two-point correlation function
$\langle\epsilon_{i}\epsilon_{j}\rangle$ and $k$ is related to the
wavelength $\lambda$ of the undulations on the random parameter
landscape by $k=1/\lambda$.$\cite{14}$ We find that the critical
disorder width  $W_{c}=4-2|E|$ at fixed energy $E$ for $\alpha>2$.
Moreover, we can determine the positions of the mobility edges $\pm
E_{c}=\pm|2-\frac{W}{2}|$, which separate the extended and localized
energy eigenstates. The effective energy band width $B$ of the
extended states has the linear relationship $B=4-W$ for $\alpha>2$.
Thus, a phase diagram in the $(E,W)$ space is obtained. We also
discuss the $\alpha$-dependence of the critical exponent $\nu$ of
the localization length of the eigenstates. Interestingly, we find
that the disorder strength $W$ determines the mobility edges and the
degree of correlation $\alpha$ determines the critical exponents.
Our results also indicate that the nature of a correlated disordered
system is somehow between that of the pure random (without any
correlation) and pseudo-random (quasi-periodic) systems.  Our
approach is non-perturbative and our results are hard to be obtained
by perturbative calculations.

The organization of the paper is as follows. In the next section, we
introduce the 1D disordered systems with long-range correlation and
the basic approach in our calculation. In section III, we present
our results of the mobility edge and critical exponents. The paper
is summarized in section IV.

\section{Model and the Basic approach}

We consider a 1D tight-binding Hamiltonian with long-range
correlated disorder
\begin{eqnarray}
H=\sum_{n}\epsilon_{n}|n\rangle\langle n|+ \sum_{n} t(
|n\rangle\langle n+1|+|n+1\rangle\langle n|),
\end{eqnarray}
where $|n\rangle$ is the Wannier state localized at site $n$ with
on-site energy $\epsilon_{n}$, and $t$ is the nearest-neighbor
hopping amplitude. We set $t=1$ for simplicity.  The on-site
energies $\{\epsilon_{n}\}$ can be constructed by the Fourier
filtering method$\cite{23,24,25}$ as
 follows: (i) generate a random sequence $\{u_{n}\}$ with a Gaussian
 distribution;
 (ii) get its Fourier components $\{u_{k}\}$ using the fast-Fourier
 transformation method; (iii) establish a new sequence $\{\epsilon_{k}\}$ by
 the relation $\epsilon_{k}=k^{-\alpha/2}u_{k}$; (iv) construct the
 sequence $\{\epsilon_{n}\}$, which is the inverse Fourier
 transform of $\{\epsilon_{k}\}$; (v) adjust the scale of the sequence
$\{\epsilon_{n}\}$ reaching to $[-W/2, W/2]$. It can be checked that
the disordered on-site energy  $\{\epsilon_{n}\}$ is long-range
correlated with the power spectrum $S(k)\propto \sum_m \langle
\epsilon_n \epsilon_{n+m} \rangle e^{imk}  \propto
k^{-\alpha}$.$\cite{26}$ The exponent $\alpha$ characterizing the
correlation reflects the roughness of the energy sequence. The
larger the value of $\alpha$ is, the smoother the energy landscape
is.

The Schr\"{o}dinger equation for the wavefunction amplitudes
$\psi_{n}$ is
\begin{eqnarray}
\epsilon_{n}\psi_{n}+\psi_{n-1}+\psi_{n+1}=E\psi_{n}.
\end{eqnarray}
Here, $E$ is the eigenenergy. Our interest focuses on the critical
behavior in the thermodynamical limit. A useful way to find the
critical points is based on the finite size analysis of the
normalized localization length $\Lambda =\xi /N$ :$\cite{26}$ If
$\Lambda $ increases as the system size $N\rightarrow\infty$, which
implies that $\xi$ would exceed $N$ when $N$ is large enough, the
electron stays on a extended state; Otherwise, the state is
localized, because $\Lambda $ is a decreasing function of $N$ and
smaller than unity in the thermodynamic limit. So the $N$ dependent
of $\Lambda =\xi /N$ can give the information of the
localization-delocalization transition. Thereby, we can use this
quantity   $\Lambda$ to determine the critical disorder strength at
fixed energy and the critical energies at fixed disorder strength.

In the regime of weak disorder, it was found$\cite{17}$ that the
localization length $\xi$ for a state of energy $E=2\cos\mu$ (not
near band center and band edge) is determined by the correlation
function of the disorder potential
\begin{eqnarray}
\xi^{-1}=\gamma_0 \varphi (\mu),
\end{eqnarray}
where $\gamma_0=\frac{\epsilon_0^2}{8 \sin^2(\mu)}$,
$\varphi(\mu)=1+2\sum_{m=1}^{\infty}\cos(2\mu m)q_m$, $\langle
\epsilon_n\epsilon_{n+m}\rangle=\epsilon_0^2 q_m$, $\langle
\epsilon_n^2 \rangle=\epsilon_0^2$. For our disordered system with
long range correlation with the spectral density $S(k)\propto
k^{-\alpha}$, the localization length has the normalized form
$\Lambda=\xi/N \sim N^{\alpha-2}$, for $ 1 \leq \alpha \leq 2$;
$\Lambda \sim 1/N$, for $\alpha < 1$. Thus we see that in the weak
disorder regime the states are localized  for $\alpha$ $<2$ and
$\alpha=2$ is a critical value for the appearance of extended
states.

To determine the full phase diagram and study the critical
parameters, we use the transfer matrix method. Using the
two-component vector $\Phi_{n}\equiv(\psi_{n},\psi_{n-1})^{T} $,
where the superscript $T$ denotes the transpose, Eq. (2) can be
written in a recursive form
\begin{eqnarray}
\left(
   \begin{array}{c}
     \psi_{n+1} \\ \psi_{n}
   \end{array}
 \right)=M_{n}\left(
   \begin{array}{c}
     \psi_{n}   \\ \psi_{n-1}
   \end{array}
 \right)
\end{eqnarray}
with
\begin{eqnarray}
M_{n}=\left(
        \begin{array}{cc}
          E-\epsilon_{n} & -1 \\
                1        &  0 \\
        \end{array}
      \right).
\end{eqnarray}
We obtain the transfer-matrix equation $\Phi_{N+1}=T_{N}\Phi_{1}$,
with $T_{N}=\prod_{i=1}^{N}M_{i}$. The localization length $\xi$ at
energy $E$ is  the inverse of the Lyapunov coefficient $(\gamma>0)$,
which is the largest eigenvalue of the limiting matrix
$\lim_{N\rightarrow\infty}\ln(T_{N}^{}T_{N}^{\dag})^{1/2N}$.$\cite{27}$
The large numbers appeared in the calculations have been took care
of by dividing a large number and multiplying it at the end of
calculation. The reorthogonalization method \cite{reorth} has also
been used and the same critical parameters are obtained. All the
values of $\gamma$ in this paper are based on a geometrical mean of
$10^{4}$ disorder configurations.

\section{Results and discussions}
\subsection{Critical disorder strength and mobility edges}
\begin{figure}[tbp]
\includegraphics*[width=0.9\linewidth]{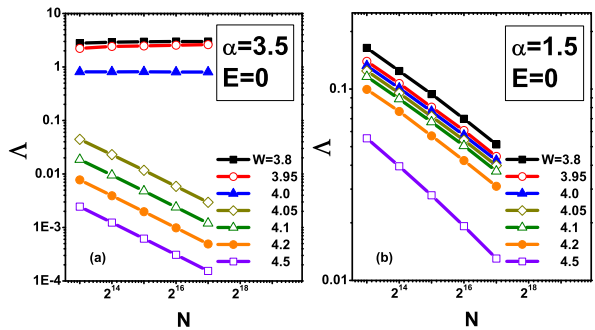}
 \caption{\small The normalized localization length $\Lambda\equiv\xi/N$
 as a function of the system size $N$ for fixed $E=0$.
 $\alpha$ is taken as $3.5$ and $1.5$ in $(a)$ and $(b)$, respectively.}
\end{figure}

We perform the calculations of the normalized localization length
$\Lambda$ for different values of $\alpha$ and energies. Figure 1
shows plots of the normalized localization length $\Lambda(W)$
versus $N$ for $E=0$, $\alpha =1.5$ and $\alpha =3.5$, respectively.
The system size $N$ ranges from $2^{13}$ to $2^{17}$. We notice that
$\Lambda$ is monotonically decreasing with $N$ for $\alpha =1.5$;
while the picture is different for $\alpha =3.5$: the values of
$\Lambda$ rise with increasing $N$ for $W<4$, whereas the values of
$\Lambda$ decline with increasing $N$ for $W>4$. The size
independence of $\Lambda$ at $W=4$ indicates a critical point of a
continuous phase transition.

In Fig. 2, we display the dependence of $\Lambda$ on the disorder
strength  $W$ for the state of $E=1$ with typical values of
$\alpha$. One may observe that there emerges an intersection point
for different system size when $\alpha=2.5$, which corresponds to
the critical point. This critical point doesn't exist for
$\alpha=1.5$. From the inset of Fig. 2, we could explicitly note the
different dependence of $\Lambda$ on the system size $N$ in the
vicinity of $W_{c}=2$. It tells us the  state with $E=1$ is
localized for $\alpha=1.5$ and is extended for $\alpha=2.5$ and
$W<2$.

\begin{figure}[tbp]
\includegraphics*[width=0.8\linewidth]{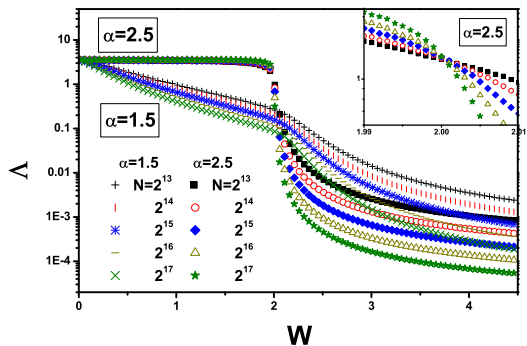}
 \caption{\small The normalized localization length $\Lambda$ as a function of
 the disorder width $W$ at fixed $E=1$ for systems with different size.
$\alpha=1.5$ and $\alpha=2.5$. Insert: The detail behavior of
$\Lambda(W)$ near the critical point $W_c=2$.}
\end{figure}

\begin{figure}[tbp]
\includegraphics*[width=0.8\linewidth]{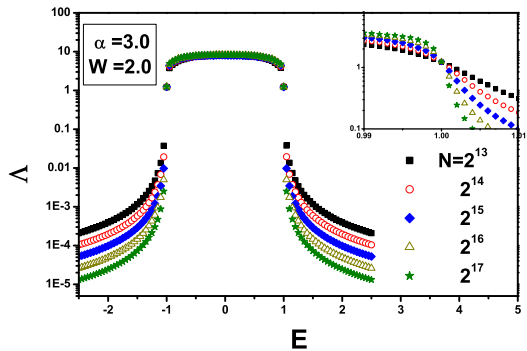}
 \caption{\small The normalized localization length $\Lambda$ as a function of
 $E$ at fixed $W=2$ for $\alpha=3.0$.
 Detail behavior of $\Lambda(E)$ near the critical point is shown in the inset.}
\end{figure}

Table I(a) lists the critical values of $W_{c}$ for various
energies. It can be seen that the critical disorder strength is
independent of $\alpha$ whenever $\alpha>2$. At the same time, we
find a simple relationship between $W_{c}$ and $E$ as:
\begin{eqnarray}
W_{c}=4-2|E|.
\end{eqnarray}
For the special state at band center with energy $(E=0)$,  the
critical value $W_c=4$, equals to the bandwidth. This conclusion for
$E=0$ state was also reached by H. Shima $\emph{et al}$.$\cite{26}$

Using the same method, we also obtain the critical energies $\pm
E_{c}$ at fixed $W$, as shown in Fig. 3. The results are given in
Table I(b) for $\alpha>2$. The critical energies/the mobility edges
can be determined by the following equations:
\begin{eqnarray}
\pm E_{c}=\pm |2-\frac{W}{2}|~~~ (\alpha>2).
\end{eqnarray}
The eigenstates between $-E_{c}$ and $E_{c}$ are delocalized.
Therefore, the effective bandwidth $B$ of the delocalized states is
(for $\alpha >2$)
\begin{eqnarray}
B=|E_{c}-(-E_{c})|=4-W ~~~(\alpha>2).
\end{eqnarray}

Figure 4 shows the phase diagrams in the $(W,E)$ and $(E,\alpha)$
plane. The critical value $\alpha=2$ is also consistent with the
perturbative calculation discussed above. We emphasize that mobility
edges or the effective bandwidth is only dependent on the disorder
strength $W$ as long as $\alpha>2$. Also, in the limit of $W=0$, the
effective bandwidth $B$ becomes $4$ which is the well-known result
in the 1D ordered crystal lattice. It is interesting to compare with
the pseudo random systems with  quasi-periodic  on-site energy
$\varepsilon_n=\frac{W}{2}\cos(\pi Q n^{\alpha}) \in
[-\frac{W}{2},\frac{W}{2}]$, Q an irrational number, $0<\alpha<1$.
In this model, the critical value $W_c=4-2|E|$ and the mobility is
also at $\pm |2- W/2|$, independent of $\alpha$.$\cite{28}$ The
linear relations between the mobility edge $E_c$ (critical disorder
strength $W_c$) and disorder strength $W$ (energy $E$) are simple
and neat, further more they have a direct consequence on the
critical exponents as will be discussed in the next subsection.

\begin{table*}

\centering {I(a) Critical values of disorder strength $W_{c}$ for
different $\alpha$ and energies $E$}

\begin{tabular}{|c|c|c|c|c|c|c|c|}\hline
  &\parbox[t]{18mm}{E=-1.5}&\parbox[t]{18mm}{E=-1.0}&\parbox[t]{18mm}{E=-0.5}&\parbox[t]{18mm}{E=0.0}&\parbox[t]{18mm}{E=0.5}&\parbox[t]{18mm}{E=1.0}&\parbox[t]{18mm}{E=1.5}\\\cline{2-8}
  \raisebox{1.5ex}[0pt]{\parbox[t]{1cm}{$\alpha$}}&$W_{c}$&$W_{c}$&$W_{c}$&$W_{c}$&$W_{c}$&$W_{c}$&$W_{c}$\\\hline
  2.50 &1.000(1)&2.001(1)&3.001(2)&4.001(2)&3.001(2)&2.001(2)&1.000(2)\\\hline
  3.00 &1.0000(4)&2.0000(5)&3.0001(5)&4.0001(6)&3.0001(6)&2.0001(6)&1.0001(4)\\\hline
  3.50 &1.0000(1)&2.0000(1)&3.0000(1)&4.0000(1)&3.0000(1)&2.0000(1)&1.0000(1)\\\hline
  4.00 &1.0000(1)&2.0000(1)&3.0000(1)&4.0000(1)&3.0000(1)&2.0000(1)&1.0000(1)\\\hline
\end{tabular}

\vskip 0.4cm

\centering{$~$} \centering {I(b) Critical values of energies $\pm
E_{c}$ for different $\alpha$ and disorder strength $W$}

\begin{tabular}{|c|c|c|c|c|c|c|}\hline
  &\multicolumn{2}{c|}{W=3.0}&\multicolumn{2}{c|}{W=2.0}&\multicolumn{2}{c|}{W=1.0}\\\cline{2-7}
  \raisebox{1.5ex}[0pt]{\parbox[t]{1cm}{$\alpha$}}&\parbox[t]{21.3mm}{$-E_{c}$}&\parbox[t]{21.3mm}{$E_{c}$}&\parbox[t]{21.3mm}{$-E_{c}$}&\parbox[t]{21.3mm}{$E_{c}$}&\parbox[t]{21.3mm}{$-E_{c}$}&\parbox[t]{21.3mm}{$E_{c}$}\\\hline
  2.50 &-0.5004(6)&0.5003(5)&-1.0001(4)&1.0002(5)&-1.5001(5)&1.5001(5)\\\hline
  3.00 &-0.5000(2)&0.5001(2)&-1.0000(3)&1.0000(2)&-1.5000(1)&1.5001(2)\\\hline
  3.50 &-0.5000(1)&0.5000(1)&-1.0000(1)&1.0000(1)&-1.5000(1)&1.5000(1)\\\hline
  4.00 &-0.5000(1)&0.5000(1)&-1.0000(1)&1.0000(1)&-1.5000(1)&1.5000(1)\\\hline
\end{tabular}

  \centering
  \caption{\small The values of critical disorder strength  $W_{c}$ and critical energies
(mobility edges)  $\pm E_{c}$ obtained by the coordinates of the
intersection points in the cases $\alpha>2$. The errors for the last
digits are listed in the parentheses.}
\end{table*}

\begin{figure}[tbp]
\includegraphics*[width=0.9\linewidth]{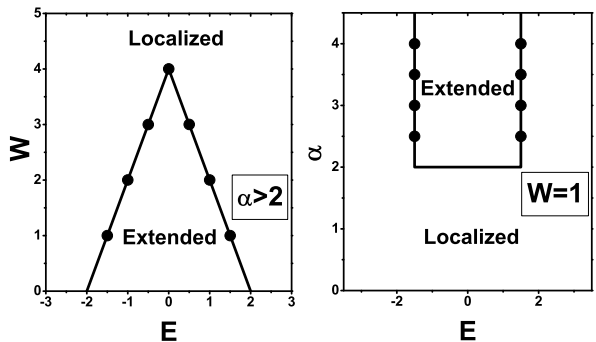}
 \caption{\small Phases diagrams in the $(W,E)$ and $(E,\alpha)$ plane.
 The delocalized and localized states are separated by $W=4-2|E|$ and $\alpha=2$.
 Here $t=1$. We take $W=1$ in $(E,\alpha)$ plane as an example.}
\end{figure}

\subsection{Critical exponents based on finite-size Scaling Analysis}
\begin{figure}[tbp]
\includegraphics*[width=0.8\linewidth]{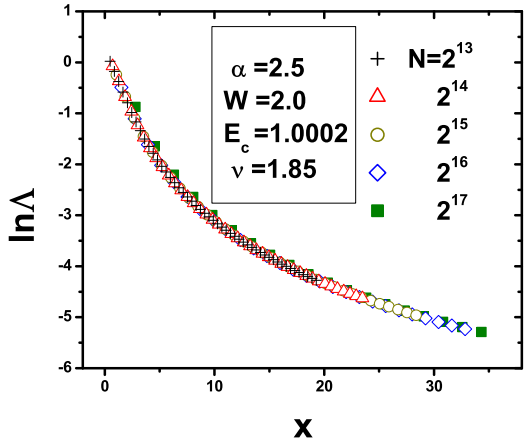}
 \caption{\small Scaling function of $\ln\Lambda$ for $\alpha=2.5$.
 Data from various chains, whose size $N$ ranges from $2^{13}$ to $2^{17}$, overlap each other.
 As an example, we give the optimized values of $E_{c}$ and $\nu$ in the case of $W=2.0$.}
\end{figure}

\begin{table*}

\centering {II(a) Critical values for disorder strength $W_{c}$ and
exponent $\nu_{W}$ for different $\alpha$ and energy $E$}

\begin{tabular}{|c|c|c|c|c|c|c|c|c|}\hline
  &\multicolumn{2}{c|}{E=0.0}&\multicolumn{2}{c|}{E=0.5}&\multicolumn{2}{c|}{E=1.0}&\multicolumn{2}{c|}{E=1.5}\\\cline{2-9}
  \raisebox{1.5ex}[0pt]{\parbox[t]{1cm}{$\alpha$}}&\parbox[t]{18mm}{$W_{c}$}&\parbox[t]{18mm}{$\nu_{W}$}&\parbox[t]{18mm}{$W_{c}$}&
  \parbox[t]{18mm}{$\nu_{W}$}&\parbox[t]{18mm}{$W_{c}$}&\parbox[t]{18mm}{$\nu_{W}$}&\parbox[t]{18mm}{$W_{c}$}&\parbox[t]{18mm}{$\nu_{W}$}\\\hline
  2.25 &4.0015(5)&2.15(5)&3.0013(3)&2.13(4)&2.0008(7)&2.12(6)&1.0002(6)&2.12(6)\\\hline
  2.50 &4.0007(5)&1.86(4)&3.0007(3)&1.83(3)&2.0004(3)&1.84(3)&1.0003(3)&1.81(3)\\\hline
  2.75 &4.0003(3)&1.65(2)&3.0003(2)&1.65(2)&2.0001(2)&1.66(2)&1.0000(2)&1.66(3)\\\hline
  3.00 &4.0002(1)&1.51(1)&3.0001(1)&1.50(1)&2.0001(1)&1.50(1)&1.0000(1)&1.50(2)\\\hline
  3.25 &4.0000(1)&1.40(1)&3.0000(1)&1.40(1)&2.0000(1)&1.40(1)&1.0000(1)&1.41(2)\\\hline
  3.50 &4.0000(1)&1.31(1)&3.0000(1)&1.30(1)&2.0000(1)&1.31(1)&1.0000(1)&1.31(1)\\\hline
  3.75 &4.0000(1)&1.24(1)&3.0000(1)&1.24(1)&2.0000(1)&1.24(1)&1.0000(1)&1.25(2)\\\hline
  4.00 &4.0000(1)&1.20(2)&3.0000(1)&1.19(1)&2.0000(1)&1.19(1)&1.0000(1)&1.20(2)\\\hline
  4.25 &4.0000(1)&1.13(2)&3.0000(1)&1.13(1)&2.0000(1)&1.13(1)&1.0000(1)&1.14(1)\\\hline
  4.50 &4.0000(1)&1.10(2)&3.0000(1)&1.10(1)&2.0000(1)&1.10(2)&1.0000(1)&1.11(2)\\\hline
\end{tabular}
\vskip 0.4cm

\centering{$~$} \centering {II(b) Critical values for energy $E_{c}$
and exponent $\nu_{E}$ for different $\alpha$ and disorder strength
$W$}

\begin{tabular}{|c|c|c|c|c|c|c|}\hline
  &\multicolumn{2}{c|}{W=3.0}&\multicolumn{2}{c|}{W=2.0}&\multicolumn{2}{c|}{W=1.0}\\\cline{2-7}
  \raisebox{1.5ex}[0pt]{\parbox[t]{1cm}{$\alpha$}}&\parbox[t]{24.5mm}{$E_{c}$}&\parbox[t]{24.5mm}{$\nu_{E}$}&\parbox[t]{24.5mm}{$E_{c}$}
  &\parbox[t]{24.5mm}{$\nu_{E}$}&\parbox[t]{24.5mm}{$E_{c}$}&\parbox[t]{24.5mm}{$\nu_{E}$}\\\hline
  2.25 &0.5007(4)&2.15(4)&1.0004(3)&2.14(3)&1.5003(4)&2.12(4)\\\hline
  2.50 &0.5003(2)&1.84(2)&1.0002(2)&1.85(2)&1.5001(2)&1.84(3)\\\hline
  2.75 &0.5001(1)&1.66(1)&1.0001(1)&1.65(1)&1.5000(1)&1.66(1)\\\hline
  3.00 &0.5001(1)&1.51(1)&1.0001(1)&1.51(1)&1.5000(1)&1.52(1)\\\hline
  3.25 &0.5000(1)&1.40(1)&1.0000(1)&1.41(1)&1.4999(1)&1.43(2)\\\hline
  3.50 &0.5000(1)&1.30(1)&1.0000(1)&1.31(1)&1.5000(1)&1.32(1)\\\hline
  3.75 &0.5000(1)&1.24(1)&1.0000(1)&1.24(1)&1.5000(1)&1.25(1)\\\hline
  4.00 &0.5000(1)&1.19(1)&1.0000(1)&1.19(1)&1.5000(1)&1.19(2)\\\hline
  4.25 &0.5000(1)&1.14(1)&1.0000(1)&1.14(1)&1.5000(1)&1.15(1)\\\hline
  4.50 &0.5000(1)&1.10(1)&1.0000(1)&1.10(2)&1.5000(1)&1.11(2)\\\hline
\end{tabular}

  \centering
  \caption{\small The values of critical exponent $\nu$ and critical
  points $W_{c}$ and $E_{c}$ got from the fitting $\Lambda $ with Eq. (10) for $\alpha>2$. The errors are obtained from bootstrap method
  with 95$\%$ confidence. }
\end{table*}

In the light of the finite-size scaling analysis, the normalized
localization length $\Lambda$ for finite system may be related to
the localization length $\xi_{\infty}$ for infinite system by the
following scaling law:$\cite{27}$
\begin{eqnarray}
\ln\Lambda=f(\frac{N}{\xi_{\infty}}),
\end{eqnarray}
where the parameter $\xi_{\infty}$ varies as
$\xi_{\infty}=|W-W_{c}|^{-\nu_{W}}$ or $\xi_{\infty}=|E\mp
E_{c}|^{-\nu_{E}}$ in the vicinity of the critical point. Unlike
three dimensional systems with usually small values of scaling
variable (i.e. the size of cross section),  in our one-dimensional
systems, the irrelevant scaling exponents \cite{ander} are of little
effect due to scaling variable with very large values, i.e., the
size of the system.

Firstly, we expand the scaling function in
polynomial form:
\begin{eqnarray}
\ln\Lambda=a_{0}+a_{1}x+a_{2}x^{2}+\cdot\cdot\cdot+a_{n}x^{n}.
\end{eqnarray}
Here $x$ defined as $x\equiv|W-W_{c}|N^{1/\nu_{W}}$ or $x\equiv|E\mp
E_{c}|N^{1/\nu_{E}}$, is a nonlinear combination of $W(E)$ and $N$
with the parameters $W_{c}(\pm E_{c})$ and $\nu$ to be determined.
Secondly, based on Eq. (10), we can obtain the values of $W_{c}(\pm
E_{c})$ and $\nu$ by fitting $\Lambda$ which we got from previous
calculation. The expansion in the scaling function must be carried
up to $n$, that makes sure the degree of confidence reaches to
$99\%$. As shown in Fig. 5, we have an excellent scaling curve,
accompanied by the optimized values of $W_{c}(\pm E_{c})$ and $\nu$.
The critical values for $W_{c}(\pm E_{c})$ obtained from finite-size
scaling analysis are identical to those obtained in section III(A).
The values of critical exponent $\nu$ and their errors estimated
from the fitting procedure are given in Table II. It is clear that
the critical exponent $\nu$ is independent on $W_{c}$ or $\pm
E_{c}$.  With increasing the correlation ($\alpha$), $\nu $
decreases to $1$. One may note that $\nu_E=\nu_W$. Actually it is a
direct consequence of the linear relation between the disorder
strength and mobility edges. From $\xi \sim |W-W_c|^{-\nu_W}$ and
Eqs. (6) and (7), one finds that $\xi \sim |E-E_c|^{-\nu_W}$. Thus
one has $\nu_E=\nu_W$. It is interesting to compare our results with
those in 3D systems. For 3D Anderson model (without correlation)
$\nu_W$ is close $\nu_E$, $\cite{16}$ while in the 3D system with
scale free disorder, $\nu_W \neq \nu_E$. $\cite{16b}$

Then we find the dependence of $\nu$ on $\alpha$ follows the rule:
\begin{eqnarray}
\nu=1+A_{1}\exp(A_{2}(2-\alpha)),
\end{eqnarray}
with $A_{1}$ is approximately $1.4$ and $A_2=1$ as seen in Fig. 6.
We have found that $\nu=1$ for infinite large correlation of
$\alpha$. It is interesting to note that the case with infinite
large long-range correlation ($\alpha \rightarrow \infty$) looks
like the quasiperiodic system, since the position of the mobility
edges and the critical exponents (1) in both systems are the same. $\cite{28}$
In another limit situation $\alpha=2$, $\nu=2.4$, which is not easy
to be obtained from direct calculation due to the critical slowing
down phenomena. This result is apparently different from $\nu=2$ at
$\alpha=2$ mentioned in Ref. $\cite{14}$. Our results are based on
finite size scaling. The data show very good scaling behavior as
shown in Fig. 5. The critical exponents and mobility edges are
obtained simultaneously by using fitting formula Eq. (10). The
mobility edges obtained in this way are identical to those obtained
in section III(A). Therefore our results are reliable. It is
interesting that the mobility edges are determined only by the
disorder strength $W$ and the critical exponents are determined only
by the degree of correlation $\alpha$.

\begin{figure}[tbp]
\includegraphics*[width=0.8\linewidth]{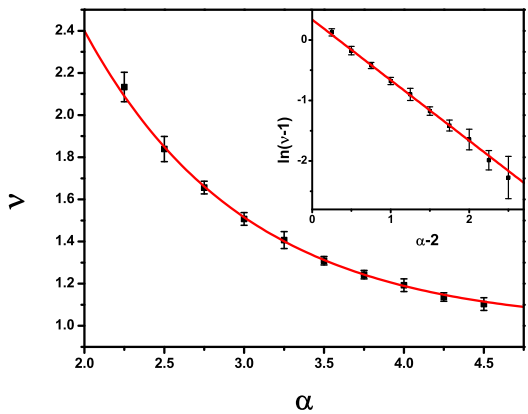}
 \caption{\small The $\alpha$-dependence of the averaged critical exponent $\nu$ (for different values of $W_c$ and $E_c$) and the fitting curve from Eq. (11)
 with $A_{1}=1.4$ and $A_{2}=1$. Inset: $\ln(\nu-1)$ as a function of $\alpha-2$, where the straight
 slope is set to $-1$.}
\end{figure}

\section{Conclusion}
In summary, we have investigated the universal localization
properties of 1D tight-binding model where the disorder is
long-range correlated with a power-law spectral density $S(k)\propto
k^{-\alpha}$, $\alpha>0$. We have obtained the complete phase
diagram and neat results for the critical parameters.  There is a
finite range of extended eigenstates, and the effective bandwidth
decreases linearly with increasing $W$ for $\alpha>2$. The positions
of the mobility edges separating localized and extended states
depend only on the disorder strength whenever $\alpha>2$. Using the
finite-size scaling analysis, we find the critical exponent
$\nu=1+1.4\exp(2-\alpha)$. In particular $\nu=2.4$ when $\alpha=2$
and $\nu=1$ when $\alpha \rightarrow \infty$. Compared to the 3D
uncorrelated Anderson model, the positions of the mobility edges are
apparently different. And the critical exponent $\nu$ is no longer
equal to $1.58$ \cite{16,ander}. So, the transition in the 1D
disordered system with correlation is in a new universal class. The
existence of mobility edge and $\alpha$ dependence of the critical
exponent indicate that the nature of a correlated disordered system
is somehow between that of the pure random (without any correlation)
and pseudo-random (quasi-periodic) systems. Our work sheds some
light on the metal-insulator transitions.

\section{Acknowledgements}
This work is supported in part by the National Natural Science of
China under No. 10574017, 10744004, 10874020, National Fundamental
Research of China grant under No. 2006CB921400 and a grant of the
China Academy of Engineering and Physics.


\end{document}